\documentclass[final]{pasj00}

\Received{$\langle$reception date$\rangle$}
\Accepted{$\langle$acception date$\rangle$}
\Published{$\langle$publication date$\rangle$}
\SetRunningHead{O. Kaburaki}{Global Operation of resistive RIAFs 
in Preparing Jet-Driving Circumstances}

\usepackage{times}

\begin{document}

\title{Global Operation of Resistive, Radiation-Inefficient, Accretion Flows \\
       in Preparing Jet-Driving Circumstances}

\author{Osamu \textsc{Kaburaki},
}
\affil{Otomaru-machi 249, Hakusan-shi, Ishikawa 924-0826
}
\email{okab@amail.plala.or.jp}


\maketitle

\begin{abstract}

In our recent paper, we have obtained a model solution to the problem of radiation-inefficient accretion 
flows (RIAFs) in a global magnetic field (so called, resistive RIAF model), which is asymptotically 
exact in outer regions of such flows forming accretion disks. When extrapolated inwardly, the model 
predicts a local enhancement of the vertical Poynting flux within a small radius that may be regarded 
as the disk inner-edge. This fact has been interpreted as the origin of power source for the astrophysical 
jets observationally well-known to be ejected from this type of accretion disks. Since the accuracy of 
the solution may become rather poor in such inner regions, however, the ground of this assertion may not 
seem to be so firm. 
In the present paper, we develop a sophisticated discussion for the appearance of jet-driving 
circumstances, based on a much more firm ground by deriving a global solution in the same situation. 
Although the new solution still has an approximate nature, it becomes exact in the limits not only of 
large radius but also of small radius. The analytic results clarify that the electrodynamic power is 
gathered by the Poynting flux, from outer main-disk region to feed the innermost part of an accretion 
disk. The injected power largely exceeds the local supply of work by the fluid motion. 

\end{abstract}

\section{Introduction}

The series of states of accretion disks called the radiatively inefficient accretion flows (RIAF) 
forms an optically-thin, under-luminous (usually radiating at a small fraction of the Eddigton luminosity 
of each central object) branch in the accretion-rate vs. surface-density diagram. Another separate branch 
exists in a more optically-thick (i.e., large surface-density) domain and continues from the 
standard-disk series to the slim-disk series, via a thermally unstable part, as the accretion rate 
increases (e.g., \cite{KFM08}). Specifically for the RIAF theories, a more detailed description can be 
found, e.g., in \citet{NM08}. 

The main efforts to take the effects of ordered magnetic fields into account in the accretion disk 
theories may be divided into two classes. In one class, the presence in the disk of only a toroidal 
field with even polarity (i.e., the polarity is the same on both upper and lower sides of the equatorial 
plane) is taken seriously. The resulting magnetic pressure is added to gas pressure to support the disk 
against the vertical component of gravity. Further, if the $\alpha$-prescription (\cite{KFM08}) with 
respect to the total pressure is adopted for a viscosity, an additional viscous extraction of angular 
momentum passing through the disk plane becomes possible. For example, the modifications of the 
standard-disk theory (e.g., \cite{BP07}) and of RIAFs (e.g., \cite{Oda12}) have been discussed, 
respectively, in relation to some controversial spectral features seen in cataclysmic variables and 
to the state transitions seen in Galactic black-hole X-ray binaries. 

In the other class, on the contrary, the presence of both poloidal and toroidal components of an 
ordered field are taken seriously. Since the toroidal component is considered to appear as a result of 
dragging of the vertical field lines by the rotational motion of the disk, its polarity reverses on 
both sides of the equatorial plane (i.e., an odd polarity). 
Thus, the toroidal component develops mainly outside the disk and vertically compresses the disk 
against gas pressure. Moreover, such a configuration makes it possible to vertically extract the 
angular momentum by the Maxwell stress. This point is essential in relation to the production 
of astrophysical jets (e.g., \cite{BP82}; \cite{PN83}; \cite{LPP94}) often observed to emanate from 
the vicinity of the disk inner edge. 

In most of the analytic models that is addressed to the formation of jets, however, the magnetic 
field is not treated self-consistently with the fluid disk. Self-consistent inclusion of an ordered 
magnetic field into RIAF states has been performed in a series of works by the present author (for a 
review, see \cite{Kab12}; hereafter referred to as Paper I). In this model, a twisted magnetic field 
works to extract angular momentum from the disk plasma, and the resistive dissipation converts the 
available gravitational energy into heat. This makes a good contrast with the usual RIAF models, in 
which only turbulent magnetic fields are included, and the fluid viscosity plays an essential role in 
converting energy and extracting angular momentum. Therefore, we call the former the resistive-RIAF 
model, distinguished from the latter, the viscous-RIAF model. 

It should be mentioned also that there is another series of studies in which the presence of an 
ordered magnetic field is treated self-consistently (for a review, see \cite{Frr08}). Although its 
relation to RIAFs is not so clear, Ferreira and his coworkers discuss an inner region of the accretion 
disk threaded by a global magnetic field. Their main interest is in the local (i.e., at a given radius) 
mechanisms to launch magnetohydrodynamic (MHD) jets, and the details of vertical transport of energy 
and angular momentum are investigated. 
On the other hand, the present concern of the resistive-RAIF model is to show how the energy 
can be supplied to the jet launching site from wide aria of an accretion disk. 

This paper is a direct follow-up of Paper I that has been devoted to discuss the appearance of 
the Poynting flux near the inner edge of a resistive-RIAF, which may lead to the jet launching. However, 
the discussion was based on the inward extrapolation of an outer asymptotic solution whose accuracy 
is not necessarily guaranteed in the inner region. Moreover, the outer solution has been derived by 
assuming a specific condition, which we call hereafter the extended RIAF condition (equation [9] 
in Paper I or [\ref{eqn:exRIAF}] below). This condition may seem rather arbitrary or artificial. 
Therefore, we give it up in the present paper. Instead, according to the spirit of this condition, 
we first obtain several asymptotic solutions in the outer region of an accretion disk, which are 
equivalent to each other within the accuracy to the first order in the smallness parameter $\epsilon_0$ 
(the definition will be given in the next section). 

Under the above situation, the criterion to sift a specific solution from others would be the 
wideness of its applicability range. 
Thus, we are naturally led to examine the behavior of these outer solutions in the opposite limit of small 
radius, and find that only one case among them becomes exact also in this limit. Namely, the selected one 
becomes accurate not only in the limit of large radius but also in that of small radius. 
Therefore, it may be called a global solution, although it is still an approximate one at middle radii. 
This finding is indeed a great improvement since we can discuss global operation of such accretion flows 
based on this much secure ground than before. Another advantage of this improved solution is that the 
expressions for all relevant physical quantities are written analytically in closed forms. 

The organization of this paper is as follows. In section 2, the variable-separated version of the 
governing equations are summarized, extracted from Paper I for convenience. As plausible examples of 
asymptotic solutions at large radii, four possibilities are derived in section 3 without employing the 
extended RIAF condition. By examining the behavior of these outer asymptotic solutions in the opposite 
limit of small radius, we find in section 4 that there is one and only one case in which the same 
expressions become asymptotically exact also in this limit. Full expressions for the relevant quantities 
within this global solution are derived also in this section. Using these expressions, we calculate and 
discuss in section 5 the local energy budgets of a few types. As summarized in the final section, the 
obtained results clearly show the appearance of preferable circumstances for jet launching.  

\section{Fundamental Equations}

As for notation, we completely follow Paper I, and hence adopt spherical polar coordinates, 
($r$, $\theta$, $\varphi$). The normalized versions of the radius and co-latitude are, 
$\xi\equiv r/r_{\rm in}$, and $\eta \equiv (\theta-\pi/2)/\Delta$, respectively. Here, $r_{\rm in}$ 
denotes the inner-edge radius of the accretion disk. 
All physical quantities have been expressed in the variable-separated forms, equations (16)-(30) 
in Paper I
\footnote{There is a typographic error in these equations. On the right-hand side of equation (21), 
$\tanh \eta$ should read $\tanh^2\eta$.}, 
within the geometrically-thin disk approximation where the half opening-angle of the 
accretion disk is very small, i.e., $\Delta \ll 1$. The fundamental equations for the radial-part 
functions in quasi-stationary problems (for the definition, see Paper I) are summarized below. 

\medskip
\noindent
{\bf leading equations}
\begin{itemize}
 \item magnetic flux conservation:
  \begin{equation}
  \frac{\tilde{b}_{\theta}}{\tilde{b}_r} = \frac{r}{2}\frac{d}{dr}\ln(r^2\tilde{b}_r) 
  = \frac{\xi}{2}\frac{d}{d\xi}\ln(\xi^2\tilde{b}_r).
  \label{eqn:fcont}
  \end{equation}
 \medskip
 \item Amp\`{e}re's law: 
  \begin{equation}
    \tilde{j}_r = \frac{c\tilde{b}_{\varphi}}{4\pi r}, 
  \label{eqn:jr}
  \end{equation}
  \begin{equation}
    \tilde{j}_{\theta} = \frac{c}{4\pi r}\ \frac{d}{dr}(r\tilde{b}_{\varphi}),
  \label{eqn:jth}
  \end{equation}
  \begin{equation}
    \tilde{j}_{\varphi} = \frac{c\tilde{b}_r}{4\pi r}, 
  \label{eqn:jph}
  \end{equation}
 \medskip

 \item Ohm's law: 
  \begin{equation}
    \tilde{E}_r = -\frac{1}{c}\ \tilde{v}_r\tilde{b}_{\theta}  
        \left( \frac{\Re_t}{\Re_p} - \frac{\tilde{v}_{\varphi}}{\tilde{v}_r} \right), 
  \label{eqn:ErM}
  \end{equation}
  \begin{equation}
    \tilde{E}_{\theta} = - \frac{1}{c}\ \tilde{v}_r\tilde{b}_r 
        \left( \Re_t - \frac{\tilde{v}_{\varphi}}{\tilde{v}_r} \right), 
  \label{eqn:EthM}
  \end{equation}
  \begin{equation}
    \tilde{E}_{\varphi} = - \frac{1}{c}\ \tilde{v}_r\tilde{b}_{\theta} \left( 1 - \Re_p^{-1} \right), 
  \label{eqn:EphM}
  \end{equation}
where 
  \begin{equation}
   \Re_p \equiv \frac{4\pi\Delta^2\sigma r\tilde{v}_r}{c^2}\ 
      \frac{\tilde{b}_{\theta}}{\tilde{b}_r}
   = \frac{4}{2n+1}\ \xi\ \frac{\tilde{v}_r}{v_{\rm in}}\ \frac{\tilde{b}_{\theta}}{\tilde{b}_r}, 
  \label{eqn:Rep}
  \end{equation}
  \begin{equation}
   \Re_t \equiv \frac{\tilde{b}_{\varphi}}{\tilde{b}_r}. 
  \label{eqn:Ret}
  \end{equation}
 \medskip

 \item mass continuity:
 \begin{equation}
  \frac{\tilde{v}_{\theta}}{\tilde{v}_r} = r\frac{d}{dr}\ln(r^2\tilde{\rho}\tilde{v}_r) 
  = \xi\frac{d}{d\xi}\ln(\xi^2\tilde{\rho}\tilde{v}_r). 
  \label{eqn:mcont}
 \end{equation}
 \medskip

 \item equation of motion:
  \begin{eqnarray}
     \lefteqn{ \left[ r\frac{d}{dr} \ln\tilde{v}_r 
       + D^{-1}\left\{ \Re_{\rm t}^2\ r\frac{d}{dr} \ln\tilde{b}_{\varphi}\right.\right. } \nonumber \\
       & & \qquad\qquad \left.\left. + \frac{1}{2}\ r\frac{d}{dr} \ln\left(\frac{\tilde{b}_r}{r^2}\right) 
         \right\}\right] \tilde{v}_r^2 = \tilde{v}_{\varphi}^2 - v_{\rm K}^2, 
  \label{eqn:EOMr}
  \end{eqnarray}
\begin{equation}
   \tilde{p} = \frac{1}{8\pi}\left(\tilde{b}_r^2 + \tilde{b}_{\varphi}^2\right). 
  \label{eqn:EOMth}
\end{equation}
\begin{equation}
  \Re_t = 2D \left\{r\frac{d}{dr}\ln(r\tilde{v}_{\varphi})\Bigm/ 
             r\frac{d}{dr}\ln(r^2\tilde{b}_r) \right\} 
          \ \frac{\tilde{v}_{\varphi}}{\tilde{v}_r}.  
  \label{eqn:EOMph}
\end{equation}
where 
 \begin{equation}
   D \equiv \frac{4\pi\tilde{\rho}\tilde{v}_r^2}{\tilde{b}_r^2}.  
   \label{eqn:defD}
 \end{equation}
 \medskip

 \item equation of state: 
  \begin{equation}
     \tilde{p} = K\tilde{\rho}\tilde{T}, 
 \label{eqn:EOS}
  \end{equation}
\end{itemize}
 \medskip

\noindent
{\bf subsidiary equations}
\begin{itemize}
 \item Faraday's law: 
  \begin{equation}
   \frac{1}{c}\ \frac{\partial \tilde{b}_r}{\partial t} 
   = -2\frac{\tilde{E}_{\varphi}}{r}, 
  \label{eqn:Fdr}
  \end{equation}
  \begin{equation}
   \frac{1}{c}\ \frac{\partial \tilde{b}_{\theta}}{\partial t} 
   = -\frac{1}{r}\ \frac{d}{dr}(r\tilde{E}_{\varphi}), 
  \label{eqn:Fdth}
  \end{equation}
  \begin{equation}
   \frac{1}{c}\ \frac{\partial \tilde{b}_{\varphi}}{\partial t} 
   = -\frac{1}{r}\ \left\{ \frac{d}{dr}(r\tilde{E}_{\theta}) 
      - 2 \tilde{E}_r \right\}, 
  \label{eqn:Fdph}
  \end{equation}
  \medskip

 \item charge density: 
  \begin{equation}
    \tilde{q} = \frac{\tilde{E}_{\theta}}{4\pi r}. 
  \label{eqn:chg}
  \end{equation}
  \medskip
\end{itemize}

One of the characteristic aspects of our treatment is that Ohm's law is used directly, without 
substituted into Faraday's law. 
The poloidal and toroidal magnetic-Reynolds numbers, equations (\ref{eqn:Rep}) and (\ref{eqn:Ret}), 
have been introduced in rewriting Ohm's law. The final expression for $\Re_{\rm p}$ can be 
derived by using equation (73) of Paper I. 

The expression (\ref{eqn:EOMr}) for the radial component 
of equation of motion (EOM) follows from equation (93) of Paper I, but the second term on the right-hand 
side (RHS) of the latter has been shifted to the left-hand side (LHS). The derivation may become evident, 
when one is referred to the identity 
\begin{equation}
  \frac{1}{\tilde{\rho}}\ \frac{d}{dr}\left(\frac{\tilde{b}_{\varphi}^2}{8\pi}\right) 
  = \frac{\Re_{\rm t}^2}{D}\left\{\frac{d}{dr}\ln\tilde{b}_{\varphi}\right\} \tilde{v}_r^2. 
 \label{eqn:id_1}
\end{equation}
In equation (\ref{eqn:EOMr}), $v_{\rm K}$ denotes the Kepler velocity $(GM/r)^{1/2}$ with $G$ and $M$ 
being the gravitational constant and the mass of a central object, respectively. The expression 
(\ref{eqn:EOMph}) for the azimuthal component of EOM has been quoted from equation (102) of Paper I. 

We often refer to the expansion parameter 
$\epsilon_0\equiv (\tilde{v}_r/\tilde{v}_{\varphi})_{\rm VPF}^2=\xi^{-1}$ that becomes very small at 
large radii, where $\tilde{v}_r$ and $\tilde{v}_{\varphi}$ are the radial and azimuthal components of 
the velocity, respectively. The suffix VPF means that the quantity is evaluated by the lowest-order 
solution, the vanishing Poynting-flux (VPF) solution (see Paper I for details). 

\section{Asymptotic Solutions at Large Radii}

\subsection{Removal of Extended-RIAF Condition}

The original RIAF condition, 
   \begin{equation}
     \alpha \equiv -\frac{1}{\tilde{\rho}}\ \frac{d\tilde{p}}{dr}\Bigm/\frac{GM}{r^2}
     = -\frac{r}{\tilde{\rho}}\ \frac{d\tilde{p}}{dr}\Bigm/v_{\rm K}^2 = \mbox{const.}, 
   \label{eqn:orRIAF}
   \end{equation}
has been assumed in the asymptotic region at large radii. This condition means that the ratio of 
the pressure gradient force to the gravity should be a constant in this region, which guarantees 
a characteristic nature of optically thin RIAFs, i.e., a virial-like temperature. Combined with the 
equation (68) of Paper I, which is the lowest-order version of the $r$-component of EOM in the power 
series of $\epsilon_0$, we have derived a lowest-order solution (\cite{Kab00}) in the asymptotic 
region. Since the Poynting flux vanishes identically in this solution (i.e., the VPF solution), 
it cannot explain the jet launching which is commonly expected for the accretion disks of the 
resistive-RIAF type. This is because the electrodynamic launching surely requires the supply of 
jet-driving power through the Poynting flux. 

In order to overcome this difficulty, we tried in Paper I to improve the accuracy of the solution by 
taking the first-order corrections in $\epsilon_0$ into account. In this connection, we have replaced 
the original RIAF condition by the extended RIAF condition, 
   \begin{equation}
     -\frac{r}{\tilde{\rho}}\ \frac{d}{dr}\left(\frac{\tilde{b}_{\varphi}^2}{8\pi}\right)
       \Bigm/\tilde{v}_{\varphi}^2 = \frac{\alpha}{1-\alpha}, \quad \alpha=\frac{2}{3}(1-n).  
   \label{eqn:exRIAF}
   \end {equation}
Here, $n$ is a constant which controls the radial profiles of relevant physical quantities in the VPF 
solution, and in this sense plays a similar role to the polytropic index that replaces the energy 
transfer equation (see Paper I). 
In contrast to the original RIAF condition, the extended one does not have firm grounds to stand on, 
except that it becomes identical with the original one in the VPF limit. Therefore, we cannot reject 
the criticism that it is only a makeshift policy, and if possible, such an obscure postulate should be 
avoided in deriving higher-order solutions. 

When the above postulate has been removed, the only remaining requirement is that any improved solution 
should coincide with the VPF solution in the limit of vanishing corrections. Under such circumstances, 
the new policy we adopt here is to portion out $\Re_{\rm t}$ and $D$ from equation (\ref{eqn:EOMph}) 
so as to reproduce a term that is proportional to $v_{\rm K}^2$ or $\tilde{v}_{\varphi}^2$, as the 
leading contribution from the partial-pressure gradient force appearing in the identity (\ref{eqn:id_1}). 
This requirement is equivalent to the condition (\ref{eqn:orRIAF}) or (\ref{eqn:exRIAF}), 
respectively, as far as the leading-order terms are concerned. Even if we follow this policy, the 
solution is not determined uniquely, and anyway the process of finding solutions becomes necessarily 
a kind of trial and error. 

In the following subsections, we show four examples of successful trials. They are all different 
from the outer asymptotic solution obtained in Paper I, which has been derived under the restriction 
of extended RIAF condition. Although only one of them leads to a final global solution, we dare to 
mention all of the new results. We believe that such a description will be helpful for the reader to 
become familiar with subtle insight into the strategies in finding solutions and to experience how the 
type of an accretion flow (i.e., sub- or trans-critical infall) is actually determined. If that had 
been omitted, the description of this paper would become very abrupt and less understandable. 

\subsection{Following Paper I}

In specifying the radial profiles of the velocity and magnetic fields, it is natural to first follow 
the results obtained in Paper I. The velocity components have been written as  
\begin{equation}
  \tilde{v}_{\varphi}(\xi) 
    = v_{\rm in}\ \xi^{-1/2}e^{-A\xi^{-1}}, 
   \label{eqn:Vph-1}
\end{equation}
\begin{equation}
   \tilde{v}_r(\xi) = v_{\rm in}\ \xi^{-1/2} \left(\frac{1-e^{-2A\xi^{-1}}}{2A}\right)^{1/2}, 
   \label{eqn:Vr_1}
\end{equation}
where $A$ is a positive constant whose value is specified in the course of discussion, and 
\begin{equation}
  v_{\rm in} \equiv \left(\frac{2n+1}{3}\right)^{1/2}V_{\rm K,in}, \quad 
  V_{\rm K,in} \equiv \sqrt{\frac{GM}{r_{\rm in}}}. 
 \label{eqn:defV}
\end{equation}
The profile of the $r$-component for the magnetic field has been fixed as 
\begin{equation}
  \tilde{b}_r(\xi) = B_{\rm in}\ \xi^{-(3/2-n)}e^{-(2n+1)A\xi^{-1}}. 
 \label{eqn:br_1}
\end{equation}
At the disk outer edge $\xi_{\rm out}$, the magnitude of this component becomes comparable to the 
externally imposed uniform field, $B_0$, as guaranteed by the relation 
$B_{\rm in}\equiv B_0\ \xi_{\rm out}^{3/2-n}$. 

The above specifications result in the derivatives 
\begin{eqnarray*}
  & & r\frac{d}{dr}\ln(r\tilde{v}_{\varphi}) = \frac{1}{2}\left(1+2A\xi^{-1}\right), \\
  & & r\frac{d}{dr}\ln(r^2\tilde{b}_r) = \frac{2n+1}{2}\left(1+2A\xi^{-1}\right), 
\end{eqnarray*}
and further from equations (\ref{eqn:fcont}), (\ref{eqn:Rep}) and (\ref{eqn:EOMph}), respectively, 
\begin{eqnarray}
  \lefteqn{ \tilde{b}_{\theta}(\xi) = \frac{2n+1}{4}\ B_{\rm in}(1+2A\xi^{-1})\times } \nonumber \\
    & & \qquad\qquad\qquad\qquad \times \xi^{-(3/2-n)}e^{-(2n+1)A\xi^{-1}}, 
 \label{eqn:bth_1}
\end{eqnarray}
\begin{equation}
  \Re_{\rm p} = \xi^{1/2}(1+2A\xi^{-1}) \left(\frac{1-e^{-2A\xi^{-1}}}{2A}\right)^{1/2}, 
 \label{eqn:Rep_1}
\end{equation}
and 
\begin{equation}
  \Re_{\rm t} = \frac{2D}{2n+1}\ \left(\frac{1-e^{-2A\xi^{-1}}}{2A}\right)^{-1/2}e^{-A\xi^{-1}}. 
 \label{eqn:Ret_1}
\end{equation}
It is evident that the above set of selections for $\tilde{v}_{\varphi}$ and $\tilde{b}_r$ is effective 
in keeping the expression (\ref{eqn:Ret_1}) rather simple. 

{\bf Case 1}. When we portion out $D$ and $\Re_{\rm t}$ from equation (\ref{eqn:Ret_1}) as 
\begin{equation}
  D = \frac{2n+1}{2} = \mbox{const.} \quad (1/4<D<1), 
\end{equation}
\begin{equation}
  \Re_{\rm t} = \left(\frac{1-e^{-2A\xi^{-1}}}{2A}\right)^{-1/2}e^{-A\xi^{-1}} 
  = \frac{\tilde{v}_{\varphi}}{\tilde{v}_r}, 
\end{equation}
we obtain the relation 
\begin{displaymath}
  \frac{\Re_{\rm t}^2}{D}\ \tilde{v}_r^2 = \frac{2}{2n+1}\ v_{\rm in}^2\ \xi^{-1}e^{-2A\xi^{-1}} 
  = \frac{2}{2n+1}\ \tilde{v}_{\varphi}^2, 
\end{displaymath}
reproducing one of our aimed form. 
Since the above determined $D$ is a constant smaller than unity for any value of $n$ in the allowed 
range, $-1/4<n<1/2$ (see \cite{Kab01}), the accretion flow in this case is said to be a `sub-critical'
\footnote{We follow here the terminology of the traditional accretion and wind theories (e.g., 
\cite{Bnd52}; \cite{Prk60}) in which there is a critical point in the governing differential equation 
at $D=1$ (i.e., $\tilde{v}_r=V_{\rm A}$). Actually in our treatment, however, no such criticality 
exists at $D=1$.} 
infall. Although $\tilde{E}_{\theta}=0$ as confirmed from $\Re_{\rm t}=\tilde{v}_{\varphi}/\tilde{v}_r$, 
the solution obtained in Case 1 does not belong to a VPF solution since $\Re_{\rm p}\neq 1$ (i.e., 
$\tilde{E}_r\neq 0$). 

The form of $\tilde{b}_{\varphi}$ follows from the definition of $\Re_{\rm t}$: 
\begin{eqnarray}
  \lefteqn{ \tilde{b}_{\varphi}(\xi) = \Re_{\rm t}\tilde{b}_r = B_{\rm in}\ \xi^{-(3/2-n)} \times } 
  \nonumber\\
  & & \qquad\qquad\quad \times \left(\frac{1-e^{-2A\xi^{-1}}}{2A}\right)^{-1/2}e^{-2(n+1)A\xi^{-1}}, 
 \label{eqn:bph_1}
\end{eqnarray}
which yields 
\begin{eqnarray}
  \lefteqn{ r\frac{d}{dr}\ln\tilde{b}_{\varphi} = -(1-n) } \nonumber\\
  & & \quad +2(n+1)A\xi^{-1} - \frac{1}{2}\ \frac{1-(1+2A\xi^{-1})e^{-2A\xi^{-1}}}{1-e^{-2A\xi^{-1}}}. 
 \label{eqn:LNbp_1}
\end{eqnarray}
At large radii, the first term on the RHS of equation (\ref{eqn:LNbp_1}) is the quantity of ${\cal O}(1)$ 
while the second and third terms are of ${\cal O}(\xi^{-1})$. They are referred to, respectively, 
as $\{r(d\ln\tilde{b}_{\varphi}/dr)\}_{\rm M}$ and $\{r(d\ln\tilde{b}_{\varphi}/dr)\}_{\rm R}$. 

Then, shifting only the leading-order term resulting from (\ref{eqn:LNbp_1}) to the RHS in the 
$r$-component of EOM (\ref{eqn:EOMr}), we have 
\begin{eqnarray}
  \mbox{RHS} &=& \tilde{v}_{\varphi}^2 - v_{\rm K}^2 
   - \left\{r\frac{d}{dr}\ln\tilde{b}_{\varphi}\right\}_{\rm M}\frac{\Re_{\rm t}^2}{D}\ \tilde{v}_r^2
  \nonumber \\
  &=& -V_{\rm K,in}^2\xi^{-1}\left(1-e^{-2A\xi^{-1}}\right) = -\frac{6A}{2n+1}\ \tilde{v}_r^2. 
\end{eqnarray} 
On the other hand, the LHS of EOM (\ref{eqn:EOMr}) becomes 
\begin{eqnarray}
  \mbox{LHS} &=& \left[ r\frac{d}{dr}\ln\tilde{v}_r 
      + \left\{r\frac{d}{dr}\ln\tilde{b}_{\varphi}\right\}_{\rm R}\frac{\Re_{\rm t}^2}{D} \right. 
  \nonumber \\ & & \qquad\qquad\qquad\qquad 
    \left. + \frac{1}{2D}\ r\frac{d}{dr}\ln\left(\frac{\tilde{b}_r}{r^2}\right) \right]\ \tilde{v}_r^2
  \nonumber \\
  &\simeq& -\left\{\frac{2n+9}{2(2n+1)} - 2A\right\} \tilde{v}_r^2. 
\end{eqnarray} 
In the final line of the this equation, only the lowest-order terms in $\xi^{-1}$ have been kept in the 
curly brackets in order to match the form of the RHS. Finally, equating both sides, we can fix 
the value of $A$ as 
\begin{equation}
  A = \frac{2n+9}{8(n+2)}. 
\end{equation}

{\bf Case 2}. The other option for portioning out of $D$ and $\Re_{\rm t}$ is 
\begin{equation}
  D = \frac{2n+1}{2}\ e^{2A\xi^{-1}}, 
\end{equation}
\begin{equation}
  \Re_{\rm t} = \left(\frac{1-e^{-2A\xi^{-1}}}{2A}\right)^{-1/2}e^{A\xi^{-1}}, 
\end{equation}
in which we have $\tilde{b}_{\varphi}$ of the form  
\begin{eqnarray}
  \lefteqn{ \tilde{b}_{\varphi}(\xi) = \Re_{\rm t}\tilde{b}_r = B_{\rm in}\ \xi^{-(3/2-n)} \times } 
  \nonumber\\
  & & \qquad\qquad\quad \times \left(\frac{1-e^{-2A\xi^{-1}}}{2A}\right)^{-1/2}e^{-2nA\xi^{-1}}, 
\end{eqnarray}
and the relations 
\begin{displaymath}
  \frac{\Re_{\rm t}^2}{D}\ \tilde{v}_r^2 = \frac{2}{2n+1}\ v_{\rm in}^2\xi^{-1} 
  = \frac{2}{3}\ v_{\rm K}^2, 
\end{displaymath}
\begin{eqnarray*}
  \lefteqn{ r\frac{d}{dr}\ln\tilde{b}_{\varphi} = -(1-n) } \\
  & & \quad +2nA\xi^{-1} - \frac{1}{2}\ \frac{1-(1+2A\xi^{-1})e^{-2A\xi^{-1}}}{1-e^{-2A\xi^{-1}}}. 
\end{eqnarray*}
Differently from Case1, the accretion flow in this case is a `trans-critical' (see the footnote 2) infall 
because it starts at a subcritical velocity (i.e., $D<1$) at the disk outer edge $\xi_{\rm out}$ 
($\gg1$) and reaches a super-critical value (i.e., $D>1$) at small radii ($\xi\ll 1$). 

After shifting only the term including $\{r(d\ln\tilde{b}_{\varphi}/dr)\}_{\rm M}$ to the RHS of equation 
(\ref{eqn:EOMr}), we have 
\begin{eqnarray}
  \lefteqn{ \mbox{RHS} = \tilde{v}_{\varphi}^2 - v_{\rm K}^2 
         + \frac{2(1-n)}{2n+1}\ v_{\rm in}^2\ \xi^{-1} } \nonumber\\
  & & \quad = -\frac{2n+1}{3}\ V_{\rm K,in}^2\ \xi^{-1}\left(1-e^{-2A\xi^{-1}}\right) 
      = -2A\ \tilde{v}_r^2. 
\end{eqnarray} 
The remaining LHS becomes 
\begin{equation}
  \mbox{LHS} \simeq -\left\{\frac{2n+9}{2(2n+1)} - 2A\ \frac{2n-1}{2n+1}\right\} \tilde{v}_r^2, 
\end{equation}
according to the same procedure as in Case 1. Equating both sides of this equation, we obtain 
\begin{equation}
  A = \frac{2n+9}{16n}, 
\end{equation}
which requires $n>0$ since $A$ should be positive definite. 

\subsection{Other Possibilities} 

In this subsection, we discuss fairly different types of expressions for the radial profiles of the 
velocity and magnetic fields. The new guideline in specifying them is to pay special attention to the 
identity 
\begin{equation}
  \xi\frac{d}{d\xi}\ \ln F = S, \quad\mbox{or}\quad 
  \frac{dF}{d\xi} = \xi^{-1}FS,
\end{equation}
where 
\begin{eqnarray}
  F(\xi) &\equiv& \left(\frac{1-e^{-2A\xi^{-1}}}{2A}\right)^{-1}e^{-2A\xi^{-1}}, \\
  S(\xi) &\equiv& \xi^{-1}\left(\frac{1-e^{-2A\xi^{-1}}}{2A}\right)^{-1}. 
\end{eqnarray}
These are monotonic functions as shown in figure 1, and their behaviour at large radii is 
\begin{equation}
  F(\xi) \rightarrow \xi\left(1-2A\xi^{-1}\right), \quad S(\xi) \rightarrow 1 
  \quad (\mbox{as }\xi \rightarrow \infty) 
 \label{eqn:lgFS}
\end{equation}
within the accuracy to the first order in $\xi^{-1}$. 

First, the radial profile of $\tilde{b}_r$ is specified as 
\begin{equation}
  \tilde{b}_r(\xi) = B_{\rm in}\ \xi^{-2}F^{(2n+1)/2}. 
 \label{eqn:br_2}
\end{equation}
It is easy to confirm that equations (\ref{eqn:br_2}) and (\ref{eqn:br_1}) are equivalent within 
the accuracy to the first order in $\xi^{-1}$. 
Then, we have  
\begin{equation}
  \tilde{b}_{\theta}(\xi) = \frac{2n+1}{4}\ B_{\rm in}\ \xi^{-2}F^{(2n+1)/2}S, 
 \label{eqn:bth_2}
\end{equation}
from the relation 
\begin{displaymath}
  r\frac{d}{dr}\ln(r^2\tilde{b}_r) = \frac{2n+1}{2}\ S. 
\end{displaymath}

In order to keep equation (\ref{eqn:EOMph}) simple, we are naturally led to select 
\begin{equation}
  \tilde{v}_{\varphi}(\xi) = v_{\rm in}\ \xi^{-1}F^{1/2} 
      = \sqrt{\frac{2n+1}{3}}\ V_{\rm K,in}\ \xi^{-1}F^{1/2} , 
 \label{eqn:vph_2}
\end{equation}
which gives 
\begin{displaymath}
  r\frac{d}{dr}\ln(r\tilde{v}_{\varphi}) = \frac{1}{2}\ S. 
\end{displaymath}
For the radial velocity component, it turns out after some trials and errors that the specification 
\begin{equation}
  \tilde{v}_r(\xi) = v_{\rm in}\ \xi^{-1} = \sqrt{\frac{2n+1}{3}}\ V_{\rm K,in}\ \xi^{-1} , 
 \label{eqn:vr_2}
\end{equation}
and hence 
\begin{equation}
  \frac{\tilde{v}_{\varphi}}{\tilde{v}_r} = F^{1/2}, 
\end{equation}
is very interesting to examine. 
Then, the velocity ratio remains the same as in subsection 3.2, and we find very simple 
expressions for the magnetic Reynolds numbers: 
\begin{equation}
  \Re_{\rm p} = S, \quad\mbox{and}\quad  \Re_{\rm t} = \frac{2D}{2n+1}\ F^{1/2}. 
 \label{eqn:Ret_2}
\end{equation}

Similarly to the discussion in the previous subsection, there are again two possibilities in portioning 
out $D$ and $\Re_{\rm t}$ from the latter of equation (\ref{eqn:Ret_2}). 

{\bf Case 3}. The first possibility is the specifications, 
\begin{equation}
  D = \frac{2n+1}{2} = \mbox{const.} \quad (1/4<D<1), 
\end{equation}
\begin{equation}
  \Re_{\rm t} = F^{1/2} = \frac{\tilde{v}_{\varphi}}{\tilde{v}_r}, 
\end{equation}
which describe a subcritical accretion flow analogously to Case 1. In this case, we have the results 
\begin{equation}
  \tilde{b}_{\varphi}(\xi) = \Re_{\rm t}\tilde{b}_r = B_{\rm in}\ \xi^{-2} F^{n+1}, 
\end{equation}
and 
\begin{displaymath}
  \frac{\Re_{\rm t}^2}{D}\ \tilde{v}_r^2 = \frac{2}{2n+1}\ v_{\rm in}^2\ \xi^{-2}F 
  = \frac{2}{2n+1}\ \tilde{v}_{\varphi}^2, 
\end{displaymath}
\begin{displaymath}
  r\frac{d}{dr}\ln\tilde{b}_{\varphi} = -2 + (n+1)S. 
\end{displaymath}
In the last equation derived above, the two terms on the RHS of the logarithmic derivative 
of $\tilde{b}_{\varphi}$ are both of order unity at large radii, since $S\rightarrow 1$ as 
$\xi\rightarrow\infty$. 

Therefore, we shift the whole term containing $r(d\ln\tilde{b}_{\varphi}/dr)$ on the LHS of EOM 
(\ref{eqn:EOMr}) to the RHS: 
\begin{eqnarray}
  \lefteqn{ \mbox{RHS} = \tilde{v}_{\varphi}^2 - v_{\rm K}^2 
   - \left\{r\frac{d}{dr}\ln\tilde{b}_{\varphi}\right\}\ \frac{\Re_{\rm t}^2}{D}\ \tilde{v}_r^2 }
  \nonumber \\
  & & = -V_{\rm K,in}^2\ \xi^{-1}\left[1-\frac{1}{3}\left\{2n+5-2(n+1)S\right\}\xi^{-1}F\right] \\
  & & \qquad \simeq -\frac{6A}{2n+1}\ \tilde{v}_r^2, \nonumber 
\end{eqnarray} 
where the last expression is the limiting form at large radii calculated with the aid of equation 
(\ref{eqn:lgFS}). Note that all the leading-order terms in the square brackets cancel out in this 
limit and we have only a first order term ($\propto\xi^{-1}$). The terms remaining on the left is 
\begin{eqnarray}
  \mbox{LHS} &=& \left[ r\frac{d}{dr}\ln\tilde{v}_r 
     + \frac{1}{2D}\ r\frac{d}{dr}\ln\left(\frac{\tilde{b}_r}{r^2}\right) \right]\ \tilde{v}_r^2 
\nonumber\\
  &=& -\frac{1}{2(2n+1)}\left\{4n+10-(2n+1)S\right\}\ \tilde{v}_r^2 \\
  &\simeq& -\frac{2n+9}{2(2n+1)}\ \tilde{v}_r^2, \nonumber 
\end{eqnarray} 
where the last expression is also the approximate form at large radii. Therefore, this equation holds 
asymptotically at large radii, as far as 
\begin{equation}
  A = \frac{2n+9}{12}. 
\end{equation}

{\bf Case 4}. The last option in our consideration is the profiles 
\begin{eqnarray}
  D &=& \frac{2n+1}{2}\ \xi F^{-1} \nonumber \\
    &=& \frac{2n+1}{2}\ \xi\left(\frac{1-e^{-2A\xi^{-1}}}{2A}\right)e^{2A\xi^{-1}}, 
 \label{eqn:D_4}
\end{eqnarray}
\begin{equation}
  \Re_{\rm t} = \xi F^{-1/2} = \xi \left(\frac{1-e^{-2A\xi^{-1}}}{2A}\right)^{1/2}e^{A\xi^{-1}}, 
\end{equation}
which describe a trans-critical accretion flow analogously to Case 2. In this case, we have the results 
\begin{equation}
  \tilde{b}_{\varphi}(\xi) = \Re_{\rm t}\tilde{b}_r = B_{\rm in}\ \xi^{-1} F^{n}, 
\end{equation}
and 
\begin{displaymath}
  \frac{\Re_{\rm t}^2}{D}\ \tilde{v}_r^2 = \frac{2}{2n+1}\ v_{\rm in}^2\ \xi^{-1} 
  = \frac{2}{3}\ v_{\rm K}^2, 
\end{displaymath}
\begin{displaymath}
  r\frac{d}{dr}\ln\tilde{b}_{\varphi} = -2 + nS. 
\end{displaymath}

As in Case 3, both terms resulting from the above logarithmic derivative of $\tilde{b}_{\varphi}$ are 
of order unity, and hence the term containing this factor in the $r$-component of EOM should be shifted 
altogether to the right. Then, the RHS becomes 
\begin{eqnarray}
  \mbox{RHS} 
  &=& -\frac{1}{3}\ V_{\rm K,in}^2\xi^{-1}\left\{1+2nS -(2n+1)\xi^{-1}F\right\}  
  \label{eqn:RHS_4} \\
  &\simeq& -2A\ \tilde{v}_r^2, \nonumber 
\end{eqnarray} 
where the last expression is the limiting form at large radii. Again, note that the leading-order 
terms in the curly brackets have been completely cancelled out in this limit. On the other hand, the 
LHS becomes 
\begin{eqnarray} 
  \mbox{LHS}  
  &=& -\frac{1}{6}\ V_{\rm K,in}^2\xi^{-2}\ \left[\ 2(2n+1) \right.  \nonumber \\
  & & \qquad\qquad\qquad \left. + \left\{8-(2n+1)S\right\}\xi^{-1}F\right] 
  \label{eqn:LHS_4} \\
  &\simeq& -\frac{2n+9}{2(2n+1)}\ \tilde{v}_r^2. \nonumber 
\end{eqnarray} 
Therefore, the equation holds in the outer asymptotic region, when 
\begin{equation}
  A = \frac{2n+9}{4(2n+1)}. 
 \label{eqn:Ala}
\end{equation}

\section{Global Solution} 

In the previous section, we have obtained four different sets of asymptotic solutions at large radii. 
Although these sets have different expressions for any one of the relevant physical quantities (and 
their components), they are equivalent within the accuracy to the first order in $\epsilon_0$ 
($=\xi^{-1}$). As far as we remain only in this outer asymptotic region, we cannot therefore judge which 
type of the accretion flows (e.g., the sub-critical or trans-critical type), is more likely to fit for 
the reality. 
For the resolution of this problem, considerations from a global point of view are needed. Thus, we are 
led to examine the behavior of the above sets in the opposite limit of small radius. Fortunately, as 
shown below, there is one and only one case (Case 4) in which the same set also serves as the asymptotic 
solution at small radii. This means that this set can be regarded as a global solution, though the 
accuracy may be somewhat poor in the middle region. 

In order to discuss the small radius limit, we note here that 
\begin{equation}
  e^{-A\xi}\rightarrow 0, \quad F\rightarrow 0, \quad S\rightarrow 2A\xi^{-1}, 
  \quad (\mbox{as } \xi\rightarrow 0 ). 
\end{equation}
In the first three cases discussed in the pervious section, the asymptotic behavior is different on 
both sides of the $r$-component of EOM. Indeed we obtain, LHS$\propto\xi^{-3}$ and RHS$\propto\xi^{-2}$ 
in Case 1; LHS$\propto\xi^{-3}$ and RHS$\propto\xi^{-2}$ in Case 2; LHS$\propto\xi^{-3}$ and 
RHS$\propto\xi^{-1}$ in Case 3. However, in Case 4, equations (\ref{eqn:LHS_4}) and (\ref{eqn:RHS_4}) 
yield, respectively, 
\begin{equation}
  \mbox{LHS}\simeq -\frac{2n+1}{3}\ V_{\rm K,in}^2\xi^{-2}, 
\end{equation}
\begin{equation}
  \mbox{RHS}\simeq -\frac{4n}{3}\ A V_{\rm K,in}^2\xi^{-2}. 
\end{equation}
Then equating both sides, we obtain 
\begin{equation}
  A = \frac{2n+1}{4n}, 
 \label{eqn:Asm}
\end{equation}
which specifies the value of $A$ in the asymptotic solution at small radii. 

If the values of $A$ in the two asymptotic regions at large and small radii (i.e., equations 
[\ref{eqn:Ala}] and [\ref{eqn:Asm}]) coincide, the asymptotic solutions match smoothly and become 
a global solution. This actually happens when 
\begin{equation}
  n = \frac{5-\sqrt{17}}{4} \simeq \frac{1}{4}, \quad 
  A = \frac{7-\sqrt{17}}{2(5-\sqrt{17})} \simeq \frac{3}{2}. 
 \label{eqn:nA}
\end{equation}
The presence of a select value of $n$ may suggest that a preferable thermodynamic circumstance is 
required for realization of the state described by our global solution. 

The global solution indicates that the infall has a trans-critical nature as seen from equation 
(\ref{eqn:D_4}). However, this should not be interpreted as a restriction on the radial profiles of infall 
velocity, since it is fixed always by equation (\ref{eqn:vr_2}) for both Cases 3 and 4. Rather, it should 
be interpreted as a restriction on the profile of the characteristic velocity 
$V_{\rm A}\equiv\tilde{b}_r/(4\pi\tilde{\rho})^{1/2}$, and hence, on those of $\tilde{b}_r$ and 
$\tilde{\rho}$. 

Other physical quantities (than $\tilde{v}_r$, $\tilde{v}_{\varphi}$, $\tilde{b}_r$, $\tilde{b}_{\theta}$, 
and $\tilde{b}_{\varphi}$) are derived straightforwardly as follows. 
It should be noted that all quantities are written in closed forms, i.e., not in the forms of truncated 
power series in $\xi^{-1}$. We obtain from equation (\ref{eqn:defD}) 
\begin{equation}
  \tilde{\rho}(\xi) = \frac{D\tilde{b}_r}{4\pi\tilde{v}_r^2} 
    = \frac{3B_{\rm in}^2}{8\pi V_{\rm K,in}^2}\ \xi^{-1}F^{2n}, 
\end{equation}
and further substituting it in equation (\ref{eqn:mcont}), 
\begin{equation}
  \tilde{v}_{\theta}(\xi) = 2n\sqrt{\frac{2n+1}{3}}\ V_{\rm K,in}\ \xi^{-1}S. 
 \label{eqn:vthF}
\end{equation}
Equation (\ref{eqn:vthF}) indicates that the generation of a wind from an accretion disk is directly 
controlled by the parameter $n$ (differently from the result in Paper I). The direction of the wind 
is upward (i.e., vertically outgoing) when $\tilde{v}_{\theta}>0$, and downward (i.e., vertically ingoing) 
when $\tilde{v}_{\theta}<0$. 
The above selected value, 
$n\sim 0.25$, means the presence of a medium-strength upward wind. It can be seen in figure 1 that, as 
far as the profiles of the velocity components are concerned, they are essentially the same as in Paper I. 

\begin{figure}

 \begin{center}
  \FigureFile(85mm, 95mm){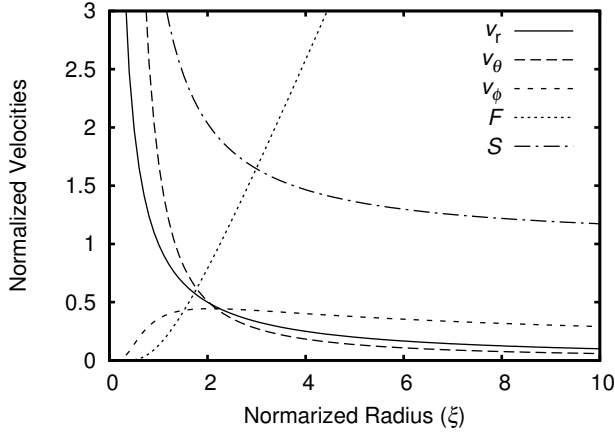}
 \end{center}
 \caption{Behavior of the radial parts ($\tilde{v}_r$, $\tilde{v}_{\theta}$, $\tilde{v}_{\varphi}$) 
of the fluid velocity are shown together with that of the functions $F(\xi)$ and $S(\xi)$. The 
velocities are normalized to $v_{\rm in}$. Although $\tilde{v}_{\theta}$ exceeds $\tilde{v}_r$ in the 
region $\xi\lesssim 2$, remember that the total expression for $v_{\theta}(\xi, \eta)$ contains a small 
factor $\Delta$. } 
\end{figure}

The pressure and temperature are calculated respectively from equations (\ref{eqn:EOMth}) and 
(\ref{eqn:EOS}) as 
\begin{equation}
  \tilde{p}(\xi) = \frac{\tilde{b}_{\varphi}^2}{8\pi}\left(1+\Re_{\rm t}^{-2}\right)
     = \frac{B_{\rm in}^2}{8\pi}\ \xi^{-2}F^{2n}\left(1+\xi^{-2}F\right), 
\end{equation}
\begin{equation}
  \tilde{T}(\xi) = \frac{\tilde{p}}{K\tilde{\rho}} 
     = \frac{1}{3}\frac{V_{\rm K,in}^2}{K}\ \xi^{-1}\left(1+\xi^{-2}F\right). 
\end{equation}
The deviation of temperature from the virial form is expected only in a middle region and remains 
to be rather small. 

The results for every component of the current density and the electric field follow from Amp\`{e}re's 
law and Ohm's law: 
\begin{equation}
  \tilde{j}_r(\xi) = \frac{cB_{\rm in}}{4\pi r_{\rm in}}\ \xi^{-2}F^n, 
\end{equation}
\begin{equation}
  \tilde{j}_{\theta}(\xi) = \frac{cnB_{\rm in}}{4\pi r_{\rm in}}\ \xi^{-2}F^n S, 
\end{equation}
\begin{equation}
  \tilde{j}_{\varphi}(\xi) = \frac{cB_{\rm in}}{4\pi r_{\rm in}}\ \xi^{-3}F^{(2n+1)/2}, 
\end{equation}
\begin{eqnarray}
  \lefteqn{ \tilde{E}_r(\xi) = -\frac{2n+1}{4}\sqrt{\frac{2n+1}{3}} 
                             \ \frac{V_{\rm K,in}B_{\rm in}}{c} \times }\nonumber \\
  & & \qquad\qquad\qquad\qquad      \times \ \xi^{-2}F^n\left(1-\xi^{-1}FS\right). 
\end{eqnarray}
\begin{equation}
  \tilde{E}_{\theta}(\xi) = -\sqrt{\frac{2n+1}{3}}\ \frac{V_{\rm K,in}B_{\rm in}}{c}
         \ \xi^{-2}F^n\left(1-\xi^{-1}F\right), 
\end{equation}
\begin{eqnarray}
  \lefteqn{ \tilde{E}_{\varphi}(\xi) = -\frac{2n+1}{4}\sqrt{\frac{2n+1}{3}} 
                             \ \frac{V_{\rm K,in}B_{\rm in}}{c} \times }\nonumber \\
  & & \qquad\qquad\qquad\qquad      \times \ \xi^{-3}F^{(2n+1)/2}\left(S-1\right). 
\end{eqnarray}
Similarly to the case of $\tilde{v}_{\theta}$, the coefficient of $\tilde{j}_{\theta}$ is proportional 
to $n$. It seems rather natural from the viewpoint of current closure that the value of $n$ determined 
in equation (\ref{eqn:nA}) is no-zero. 

Finally, we cite the component expressions for the Poynting flux, including their $\eta$-dependences 
since they have been dropped in Paper I by accident. 
\begin{eqnarray}
  \lefteqn{ P_r(\xi, \eta) = \tilde{P}_r(\xi)\ \mbox{sech}^2\eta\ \tanh^2\eta, } \nonumber \\
  & & \quad \tilde{P}_r(\xi) = \frac{c}{4\pi}\tilde{E}_{\theta}\tilde{b}_{\varphi} 
       = -\sqrt{\frac{2n+1}{3}}\ \frac{V_{\rm K,in}B_{\rm in}^2}{4\pi} \times \nonumber \\
  & & \qquad\qquad\qquad\qquad\qquad\quad \times \ \xi^{-3}F^{2n}\left(1-\xi^{-1}F\right), 
\end{eqnarray}
\begin{eqnarray}
  \lefteqn{ P_{\theta}(\xi, \eta) = \Delta\tilde{P}_{\theta}(\xi)\ \mbox{sech}^2\eta\ \tanh\eta, } 
   \nonumber \\
  & & \quad \tilde{P}_{\theta}(\xi) 
    = \frac{c}{4\pi}\left(\tilde{E}_r\tilde{b}_{\varphi}-\tilde{E}_{\varphi}\tilde{b}_r\right) 
   \nonumber\\
  & & \qquad\quad = -\frac{2n+1}{4}\ \sqrt{\frac{2n+1}{3}}\ \frac{V_{\rm K,in}B_{\rm in}^2}{4\pi} \times 
   \nonumber \\
  & & \qquad\qquad \times \ \xi^{-3}F^{2n}\left\{1-\xi^{-1}FS-\xi^{-2}F\left(S-1\right)\right\}, 
 \label{eqn:PFth}
\end{eqnarray}
\begin{eqnarray}
  \lefteqn{ P_{\varphi}(\xi, \eta) = \tilde{P}_{\varphi}(\xi)\ \mbox{sech}^4\eta\ \tanh^2\eta, } 
   \nonumber \\
  & & \quad \tilde{P}_{\varphi}(\xi) = \frac{c}{4\pi}\tilde{E}_{\theta}\tilde{b}_r 
       = -\sqrt{\frac{2n+1}{3}}\ \frac{V_{\rm K,in}B_{\rm in}^2}{4\pi} \times \nonumber \\
  & & \qquad\qquad\qquad\qquad\quad \times \ \xi^{-4}F^{(4n+1)/2}\left(1-\xi^{-1}F\right), 
\end{eqnarray}

\begin{figure}
 \begin{center}
  \FigureFile(85mm, 95mm){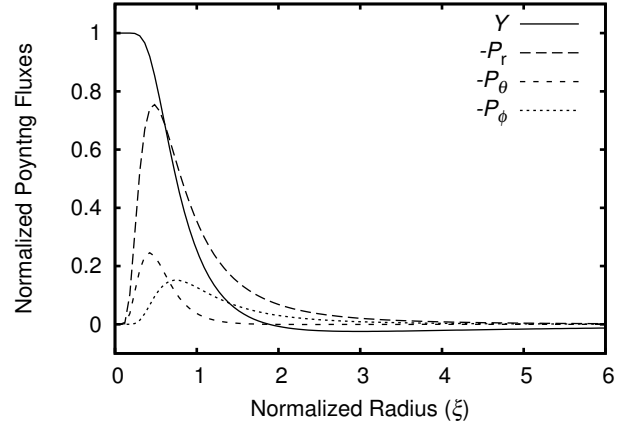}
 \end{center}
 \caption{Behavior of the sign-reversed radial functions of the Poynting flux (i.e., $-\tilde{P}_r$, 
$-\tilde{P}_{\theta}$, $-\tilde{P}_{\varphi}$). They are normalized to $v_{\rm in}B_{\rm in}^2/4\pi$. 
Although $-\tilde{P}_r$ and $-\tilde{P}_{\varphi}$ are everywhere positive, $-\tilde{P}_{\theta}$ is 
positive only where $\xi\lesssim 2$ and slightly negative where $\xi\gtrsim 2$. 
In order to see this fact more clearly, the {\bf formula enclosed by} the curly brackets in equation 
(\ref{eqn:PFth}) is shown as $Y(\xi)$. }
\end{figure}

As it turns out from figure 2, $\tilde{P}_r$ and $\tilde{P}_{\varphi}$ are negative everywhere, but 
$\tilde{P}_{\theta}$ changes sign from slightly positive (outgoing) to negative (incoming) as the 
radius decreases across $\xi\sim 2$. In order to see this fact more clearly, the formula in the curly 
brackets in equation (\ref{eqn:PFth}) is shown as $Y(\xi)$ in the figure. Taking the $\eta$-dependece 
of $P_r(\xi, \eta)$ also into account, we can say that the Poynting flux in the poloidal plane flows 
from the wide outer region ($\xi\gtrsim 2$) into the narrow inner region ($\xi\lesssim 2$), almost 
along the surface of an accretion disk. 

It should be noted that $\tilde{P}_{\theta}$ here has the opposite sign to that obtained in Paper 
I, in the inner region where $\tilde{P}_{\theta}$ becomes very large. This means that the extrapolation 
of an outer solution that is not the global solution can actually lead to an erroneous result. 

\section{Local Energy Budgets}

\subsection{Mass Accretion Rate}

Before proceeding to energy budgets, it is convenient to introduce the mass accretion rate through 
a vertical cross-section of arbitrary radius, 
\begin{eqnarray}
  \dot{M}(\xi) &\equiv& -\int_{-\infty}^{\infty}2\pi\rho v_r r^2\Delta d\eta 
     \simeq 2\pi\Delta\tilde{\rho}\tilde{v}_r r^2\int_{-\infty}^{\infty}\mbox{sech}^4\eta\ d\eta 
  \nonumber \\
  &=& \sqrt{\frac{2n+1}{3}}\ \frac{\Delta B_{\rm in}^2 r_{\rm in}^2}{V_{\rm K,in}}\ F^{2n} 
   \equiv \dot{M}_0\ \xi_{\rm out}^{-2n}F^{2n}, 
\end{eqnarray}
where the approximation $v_r = -\tilde{v}_r\mbox{sech}^2\eta$ has been used, extrapolating the 
functional form near the equator even to large $\eta$ regions (i.e., neglecting the $\tanh^2\eta$ term). 
This is because the $\eta$-dependences are reliable only near the disk midplane owing to the adopted 
method of approximation (see Paper I). 
Since we need a co-latitudinal integration in the above calculation, this may cause some worry about the 
accuracy of the result. However, the most important thing is a finiteness of the integral and its precise 
value does not matter on the essence of the following discussions. 
Reflecting the presence of the vertical flows, $\dot{M}$ varies with $\xi$ like $F^{2n}$. It is a 
constant only when $n=0$, i.e., there is no wind. 

The accretion rate at the outer edge of an accretion disk is given explicitly as 
\begin{equation}
  \dot{M}_0 \equiv \dot{M}(\xi_{\rm out}) 
    = \sqrt{\frac{2n+1}{3}}\ \frac{\Delta B_{\rm in}^2r_{\rm in}^2}{V_{\rm K,in}}\ \xi_{\rm out}^{2n}, 
\end{equation}
because, at this radius, $F$ can be approximated by its asymptotic form, $F\sim\xi$. First substituting 
$B_{\rm in}=B_0\ \xi_{\rm out}^{3/2-n}$, $V_{\rm K,in}=(GM/r_{\rm in})^{1/2}$ and 
$\xi_{\rm out}=r_{\rm out}/r_{\rm in}$ in the above definition, and then solving for $B_0^2$, we obtain 
\begin{equation}
  B_0^2 = \sqrt{\frac{3}{2n+1}}\ \frac{\sqrt{GM}\dot{M}_0}{\Delta\ r_{\rm out}^{5/2}}
     \ \xi_{\rm out}^{-1/2}. 
 \label{eqn:B0}
\end{equation}
This equation will be used to eliminate $B_0^2$ in the following subsections. 

\subsection{Thermal Energy Budget}

The heating rate due to the Joule dissipation of the electric current per unit volume is calculated as 
\begin{eqnarray}
  \lefteqn{ q_{\rm J}^{+}(\xi,\ \eta) = \frac{\mbox{\boldmath $j$}^2}{\sigma} 
            \simeq \tilde{q}_{\rm J}^{+}(\xi)\ \mbox{sech}^4\eta, }  \nonumber \\
  & & \quad \tilde{q}_{\rm J}^{+}(\xi) 
      = \frac{1}{\sigma\Delta^2}\left(\tilde{j}_r^2 + \tilde{j}_{\varphi}^2\right) \nonumber\\
  & & \qquad = \frac{2n+1}{16\pi\Delta}\ \frac{GM\dot{M}_0}{r_{\rm out}^4}\ \xi_{\rm out}^{4-2n} 
               \xi^{-4}F^{2n}\left(1+\xi^{-2}F\right). 
\end{eqnarray}
In obtaining the approximate expression in the first line of the above equations, we have neglected 
$j_{\theta}$ since $j_r\sim j_{\varphi}\sim {\cal O}(\Delta^{-1})$, $j_{\theta}\sim{\cal O}(1)$, and 
also neglected a term containing $\tanh^2\eta$ dependence. In obtaining the expression on the third 
line, $\sigma\Delta^2$ and $B_0^2$ have been eliminated with the aids of equation (73) in Paper I and 
equation (\ref{eqn:B0}) above, respectively. 

On the other hand, the advection cooling rate per unit volume is written as 
\begin{eqnarray}
  \lefteqn{ q_{\rm adv}^{-}(\xi,\ \eta) \equiv {\bf \nabla}\cdot(h\rho\mbox{\boldmath $v$}) 
                                          -(\mbox{\boldmath $v$}\cdot{\bf \nabla})\ p  
  \simeq \tilde{q}_{\rm adv}^{-}(\xi)\ \mbox{sech}^4\eta, } \nonumber \\
  & & \quad \tilde{q}_{\rm adv}^{-}(\xi) = -\frac{5}{2}\ \frac{\tilde{p}}{r^2}
      \ \frac{d}{dr}\left(r^2\tilde{v}_r\right) - \frac{3}{2}\ \tilde{v}_r\frac{d\tilde{p}}{dr} 
      + \frac{5}{2}\ \frac{\tilde{p}\tilde{v}_{\theta}}{r} \nonumber\\
  & & \qquad = \frac{1}{16\pi\Delta}\ \frac{GM\dot{M}_0}{r_{\rm out}^4}\ \xi_{\rm out}^{4-2n}
      \times \nonumber\\
  & & \qquad\qquad\quad \times\xi^{-4}F^{2n} \left\{1+4nS +(4n-3)\xi^{-2}FS\right\}, 
\end{eqnarray}
where $h=(5/2)(p/\rho)=(5/2)(\tilde{p}/\tilde{\rho})$ is the enthalpy for an ideal gas. Similarly to 
the case of $q_{\rm J}^{+}$, the approximate expression on the first line of the above equations is 
obtained by neglecting the $\tanh^2\eta$ terms. 

However, the heating and cooling terms given above does not balance generally. This is because i) there 
may exist the exchange of thermal energy between neighboring fluid elements through conduction and/or 
convection (these are expressed as non-adiabaticities), even if the radiation loss may be negligibly 
small, or ii) the balance cannot be achieved in principle, implying the break-down of the model. 
The amount of this discrepancy is 
\begin{eqnarray}
  \lefteqn{ \tilde{q}_{\rm dis}^{+}(\xi) \equiv \tilde{q}_{\rm adv}^{-}(\xi) 
            - \tilde{q}_{\rm J}^{+}(\xi) }   \nonumber \\
  & & \quad 
      = \frac{1}{16\pi\Delta}\ \frac{GM\dot{M}_0}{r_{\rm out}^4}\ \xi_{\rm out}^{4-2n}\xi^{-4}F^{2n} 
              \times \nonumber\\
  & & \qquad \times \left\{2n\left(1+\xi^{-2}F\right)(2S-1) - \xi^{-2}F(3S+1)\right\}, 
\end{eqnarray}
when it is expressed in terms of an additional heating (or a cooling if it is negative). 

There are two distinct terms in the above result; the term which contains $n$ and which does not. The 
former dominates over the latter in the outer asymptotic region (i.e., in the VPF limit). In this 
asymptotic region, this term has been interpreted as due to the non-adiabaticities (\cite{Kab01}). 
Our global solution indicates that both terms enhance in a narrow region at around $\xi\sim 0.4$.
Even in such inner regions, the former drives an upward wind (i.e., $\tilde{v}_{\theta}>0$) when there 
is additive heating (i.e., $n>0$), and vice versa, as seen in equation (\ref{eqn:vthF}). On the other 
hand, the latter term always contributes to a cooling. Although it is not so certain, this fact may 
suggests that the latter corresponds to a radiative cooling that may enhance at such an inner region. 

As shown in figure 3, the advection cooling becomes largely exceeds the Joule heating where 
$\xi\lesssim 1$. The cause of this enhancement is in the monotonic increase of $\tilde{v}_r$ and 
 $\tilde{v}_{\theta}$ toward the center. In principle, it is impossible to balance the advective cooling 
even by supplying heat through conduction or convection, as far as the cooling there exceeds the peak 
value of the Joule heating. In this sense, the present model does not seem to hold in the region where 
$\xi\lesssim 0.8$. Thus, we are consistently led to the conclusion that there is an inner edge of an 
accretion disk at $\xi\sim 1$. 

\subsection{Electromagnetic Energy Budget}

The energy equation for the electromagnetic field is known as the Poynting theorem: 
\begin{equation}
  \frac{\partial u}{\partial t} + {\bf \nabla}\cdot\mbox{\boldmath $P$} 
  = -\mbox{\boldmath $E$}\cdot\mbox{\boldmath $j$}
  = -\frac{\mbox{\boldmath $j$}^2}{\sigma} 
    -\frac{1}{c}\left(\mbox{\boldmath $j$}\times \mbox{\boldmath $b$}\right)\cdot\mbox{\boldmath $v$}, 
 \label{eqn:Pyt}
\end{equation}
where $u=(1/8\pi)(\mbox{\boldmath $b$}^2+\mbox{\boldmath $E$}^2)\simeq (1/8\pi)\mbox{\boldmath $b$}^2$ 
(see Paper I) is the electromagnetic energy density, and $\mbox{\boldmath $P$}$ is the Poynting flux. 
The two terms on the right-most side of equation (\ref{eqn:Pyt}) represent losses of the 
electromagnetic energy through the Joule dissipation and through work done by the magnetic force on 
the fluid, respectively. 

When we concentrate our attention mainly to the region near the disk midplane, we can neglect 
$\tanh^2\eta$ terms as before. The contribution from $\partial u/\partial t$ vanishes in this process, 
and the Poynting equation finally reduces to the form  
\begin{eqnarray}
  \lefteqn{ W(\xi, \eta) \equiv 
   -\frac{1}{c}\left(\mbox{\boldmath $j$}\times \mbox{\boldmath $b$}\right)\cdot\mbox{\boldmath $v$} 
  \simeq \tilde{W}(\xi)\ \mbox{sech}^4\eta, } \nonumber \\ 
  & & \quad \tilde{W}(\xi) = \tilde{q}_{\rm J}^{+} + \frac{\tilde{P}_{\theta}}{r} 
  \nonumber \\
  & & \quad = \frac{2n+1}{16\pi\Delta}\ \frac{GM\dot{M}_0}{r_{\rm out}^4}
      \ \xi_{\rm out}^{4-2n} 
\xi^{-5}\left(1+\xi^{-1}\right)F^{2n+1}S. 
\end{eqnarray}
As the above definition of $W(\xi, \eta)$ includes a minus sign, it means the work done by fluid 
on the electromagnetic force (hence, on the field). The explicit expression of $\tilde{W}(\xi)$ can 
be reached either by calculating the right-hand side of the expression on the middle line or by 
calculating its left-hand side directly according to its definition, confirming that the Poynting 
equation actually holds in our case. 

Since $\tilde{W}>0$ at any finite radius, the fluid motion is doing work on the electromagnetic 
field everywhere in the disk. As seen from figure 2, the vertical Poynting flux, $\tilde{P}_{\theta}$, 
is positive in the region $\xi\gtrsim 2$ while negative in $\xi\lesssim 2$. This implies that 
$\tilde{W}>\tilde{q}_{\rm J}^{+}$ where $\xi\gtrsim 2$ and $\tilde{W}<\tilde{q}_{\rm J}^{+}$ where 
$\xi\lesssim 2$. Namely, in the wide outer region ($\xi\gtrsim 2$) the fluid in the disk is doing work 
that exceeds the local dissipation $\tilde{q}_{\rm J}^{+}$ and the excess is gathered into the small 
inner region ($\xi\lesssim 2$) through the Poynting flux (see also the discussion at the end of the 
previous section). Thus, in the inner region, energy input through the Poynting flux can largely 
exceeds local supply through the work $\tilde{W}$, and both are thermalized as the Joule dissipation 
(i.e., $\tilde{W}-\tilde{P}_{\theta}/r=\tilde{q}_{\rm J}^{+}$). 

In other words, the outer main disk is driving the innermost region electrodynamically, 
suggesting that, if a more careful treatment of the vertical structure and flow in this region is 
introduced in the model, the launching of an MHD jet could be obtained. Within the present status 
of our accretion disk model, however, this ability is wasted only on the Joule dissipation there. 
In a sense, the new global picture of our model is reasonable because it closes within itself. 
On the other hand, the previous picture stated in Paper I is less persuasive, because it does 
not close unless assuming a presence of some external component (e.g., an MHD jet) that is driven by 
the large positive (i.e., outgoing) Poynting flux mainly emanating from the innermost disk.  

\begin{figure}
 \begin{center}
  \FigureFile(85mm, 95mm){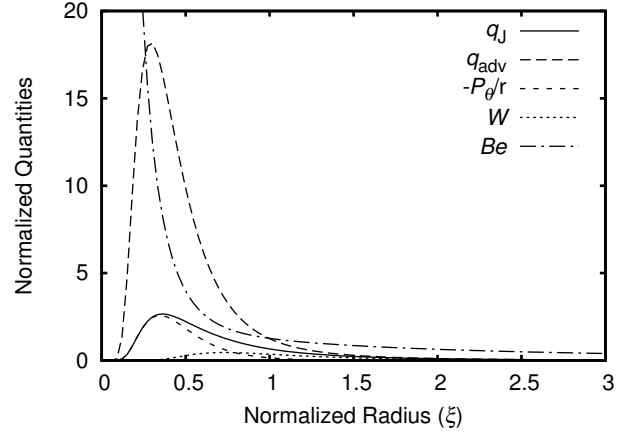}
 \end{center}
 \caption{Behaviour of $\tilde{q}_{\rm J}^{+}$, $\tilde{q}_{\rm adv}^{-}$, $-\tilde{P}_{\theta}/r$, 
$\tilde{W}$ and $\tilde{B}e$. Note that the former four quantities and the latter one have different 
dimensions, and normalized to $(GM\dot{M}_0/16\pi\Delta r_{\rm out}^4)\ \xi_{\rm out}^{(4-2n)}$ and 
$V_{\rm K,in}^2/6$, respectively. }
\end{figure}

\subsection{Binding Energy}

The binding energy per unit volume of a fluid element is described by the Bernoulli sum 
(\authorcite{NY94}\yearcite{NY94}, \yearcite{NY95}), 
\begin{eqnarray}
  B_{\rm e}(\xi,\ \eta) &\equiv& \frac{1}{2}\ \mbox{\boldmath $v$}^2 + h  - \frac{GM}{r} \nonumber\\
  &\simeq& \frac{1}{2}\left( \tilde{v}_r^2 + \tilde{v}_{\varphi}^2 \right) 
      + \frac{5}{2}\ \frac{\tilde{p}}{\tilde{\rho}} - \frac{GM}{r} 
  \equiv \tilde{B}_{\rm e}(\xi) 
\end{eqnarray}
where the approximations of geometrically-thin disk and that of respecting midplane have been used again 
in obtaining the expression on the second line. Calculating this quantity in terms of our global solution, 
we obtain 
\begin{eqnarray}
  \lefteqn{ \tilde{B}_{\rm e}(\xi) = \frac{1}{6}\ V_{\rm K,in}^2\xi^{-1}\times } \nonumber \\
     & & \qquad\qquad\quad \times \left\{(2n+1)\xi^{-1}(1+F)+5\xi^{-2}F-1\right\}. 
\end{eqnarray}

The asymptotic values for large and small $\xi$ are 
\begin{equation}
 \tilde{B}_{\rm e}(\xi) \approx \left\{
   \begin{array}{ll}
     \displaystyle{\frac{n}{3}}\ V_{\rm K,in}^2 \xi^{-1},&\quad \mbox{as }\ \xi \rightarrow \infty, \\
     \noalign{\vskip 0.2cm}
     \displaystyle{\frac{2n+1}{6}}\ V_{\rm K,in}^2 \xi^{-2},&\quad \mbox{as }\ \xi \rightarrow 0.  
   \end{array} \right. 
\end{equation}
The upper line of the above equation reproduces the VPF result (\cite{Kab01}). As seen in figure 3, 
the global solution predicts that $\tilde{B}_{\rm e}(\xi)$ remains positive everywhere 
(cf. \authorcite{NY94}\yearcite{NY94}, \yearcite{NY95}; however see also, \cite{Nak98}; \cite{ALI00}; 
\cite{BB99}; \cite{Bck00}; \cite{TD00}) and becomes divergently large in the limit of $\xi\rightarrow 0$, 
even when $n=0$. This fact suggests that the accretion flows described by our global solution should be 
ejected, at least in its fraction, before it reaches the center. 

\section{Summary}

Summarizing the discussions in the previous section, we have reached the conclusions that i) 
the accretion state characterized under the name of resistive-RIAF does not seem to extend into the 
region $\xi\lesssim 1$, ii) the main disk ($\xi\gtrsim 2$) is driving the innermost region 
($\xi\lesssim 2$) electrodynamically, and iii) infalling matter always stay unbound and cannot reach 
the gravitational center, as a whole. Judging from these evidences, the most probable scenario 
is an ejection of the infalling matter, at least in its fraction, at around $\xi\sim 1$ which may 
be regarded as the inner edge of an accretion disk. Thus, it has been shown clearly that one of 
the most preferable circumstances necessary for the MHD jet launching is actually prepared within our 
resistive-RIAF model.


\begin{thebibliography}{}

\bibitem[Abramowicz et al. (2000)]{ALI00} Abramowicz, M., Lasota, J.-P., \& Igumenshchev, I.V., 2000, 
 \mnras, 314, 775 
\bibitem[Beckert (2000)]{Bck00} Beckert, T., 2000, ApJ, 539, 223 
\bibitem[Begelman \& Pringle (2007)]{BP07} Begelman, M. C., \& Pringle, J. E., 2007, \mnras, 375, 
 1070 
\bibitem[Blandford \& Begelman (1999)]{BB99} Blandford, R. D., \& Begelman, M. C., 1999, 
 \mnras, 303, L1 
\bibitem[Blandford \& Payne (1982)]{BP82} Blandford, R. D., \& Payne, D. G., 1982, \mnras, 199, 883 
\bibitem[Bondi (1952)]{Bnd52} Bondi, H., 1952, MNRAS, 112, 195
\bibitem[Ferreira (2008)]{Frr08} Ferreira, J., 2008, New Astron. Rev., 52, 42 
\bibitem[Kaburaki (2000)]{Kab00} Kaburaki, O., 2000, ApJ, 531, 210
\bibitem[Kaburaki (2001)]{Kab01} Kaburaki, O., 2001, ApJ, 563, 505
\bibitem[Kaburaki (2012)]{Kab12} Kaburaki, O., 2012, PASJ, 64, 39, Paper I
\bibitem[Kato et al. (2008)]{KFM08} Kato, S., Fukue, J., \& Mineshige, S., 2008, 
 Black-Hole Accretion Disks: Towards a New Paradigm (Kyoto University Press, Kyoto)
\bibitem[Lubow et al. (1994)]{LPP94} Lubow, S.H., Papaloizou, J. C. B. \& Pringle, J. E. 1994, 
 \mnras, 268, 1010 
\bibitem[Nakamura (1998)]{Nak98} Nakamura, K. E., 1998, \pasj, 50, L11
\bibitem[Narayan \& McClintock (2008)]{NM08} Narayan, R. \& McClintock, J. E. 2008, New Astron. Rev. 
 51, 733 
\bibitem[Narayan \& Yi (1994)]{NY94} Narayan, R. \& Yi, I. 1994, ApJ, 428, L13 
\bibitem[Narayan \& Yi (1995)]{NY95} Narayan, R. \& Yi, I. 1995, ApJ, 444, 231 
\bibitem[Oda et al. (2012)]{Oda12} Oda, H., Machida, M., Nakamura, K. E., Matsumoto, R. \& Narayan, R. 
2012, \pasj, 64, 15 
\bibitem[Parker (1960)]{Prk60} Parker, E. N., 1960, ApJ, 132, 175
\bibitem[Pudritz \& Norman (1983)]{PN83} Pudritz, R. E. \& Norman, C. A. 1983, ApJ, 274, 677 
\bibitem[Turolla \& Dullemond (2000)]{TD00} Turolla, R. \& Dullemond, C. P. 2000, \apj, 531, L49 

\end{thebibliography}
\end{document}